\documentclass[%
 reprint,
 superscriptaddress,
 amsmath,amssymb,
 aps,
]{revtex4-2}

\usepackage{color}
\usepackage{graphicx}
\usepackage{subfigure}
\usepackage[utf8]{inputenc}
\usepackage{dcolumn}
\usepackage{bm}
\usepackage{mathrsfs}
\usepackage{tipa}

\usepackage{hyperref}

\begin{document}

\title{Measuring and Modeling Bursty Human Phenomena}\thanks{This is a draft of the chapter. The final version will be available in the Handbook of Computational Social Science edited by Taha Yasseri, forthcoming 2025, Edward Elgar Publishing Ltd. The material cannot be used for any other purpose without further permission of the publisher and is for private use only. Please cite as: M\'arton Karsai and Hang-Hyun Jo (2025). Measuring and Modeling Bursty Human Phenomena. In: T. Yasseri (Ed.), Handbook of Computational Social Science. Edward Elgar Publishing Ltd.}

\author{M\'arton Karsai}
\email{karsaim@ceu.edu}
\affiliation{Department of Network and Data Science, Central European Universtiy, Vienna, Austria}
\affiliation{National Laboratory for Health Security, HUN-REN Alfr\'ed R\'enyi Institute of Mathematics, Budapest, Hungary}

\author{Hang-Hyun Jo}
\email{h2jo@catholic.ac.kr}
\affiliation{Department of Physics, The Catholic University of Korea, Bucheon, Republic of Korea}

\date{\today}

\begin{abstract}
Bursty dynamics characterizes systems that evolve through short active periods of several events, which are separated by long periods of inactivity. Systems with such temporal heterogeneities are not only found in nature but also include examples from most aspects of human dynamics. In this Chapter, we briefly introduce such bursty phenomena by first walking through the most prominent observations of bursty human behavior. We then introduce several conventional measures that have been developed for characterizing bursty phenomena. Finally, we discuss the fundamental modeling directions proposed to understand the emergence of burstiness in human dynamics through the assumption of task prioritization, temporal correlations, and external factors. This Chapter is only a concise introduction to the field. Still, it provides the most important references, which will help interested readers to learn in depth the ever-growing research area of bursty human dynamics. 
\end{abstract}


\maketitle

\section{Introduction}
\label{sec:intro}

There are a number of natural and human dynamical systems showing similar temporal characteristics; they evolve via events appearing within short and highly active bursts that are separated by long periods of inactivity~\cite{Barabasi2005Origin, Karsai2018Bursty}. Earthquakes and solar flares are well-known examples of such bursty patterns~\cite{deArcangelis2006Universality}, as explained by the Omori's law and the theory of self-organized criticality. Firing sequences of single or a collective of neurons show similar behavior~\cite{Tsubo2012Powerlaw}. Further, the movement of animals~\cite{Boyer2012Nonrandom} or even evolution~\cite{Uyeda2011Millionyear}, regarding the time of emergence of divergent taxonomic groups, can be recognized as bursty phenomena. Meanwhile there are several examples of man-made systems, which evolve through bursty dynamics. The best examples are word occurrences in written text, package traffic in wired and wireless telecommunication systems, and the financial market~\cite{Karsai2018Bursty}.

The emerging access to large-scale digital datasets on human behavior has opened the door for detailed analysis of the actions or interactions of millions of people. Moreover, as some of these datasets were collected with no direct intervention, usually for other purposes like billing or system logs, they provide reliable observations about very large populations with weak or no observational biases. The analysis of such datasets resulted in the first convincing observations of human bursty phenomena in email sequences~\cite{Eckmann2004Entropy, Barabasi2005Origin}. Subsequently similar patterns were found in a myriad of other systems, suggesting burstiness as a universal phenomenon characterizing all kinds of aspects of human dynamics from the individual to the collective level~\cite{Karsai2018Bursty}. Actions like library loans, printing behavior, or web browsing; interactions like emailing, texting, messaging, or phone calling; and collective behavior like the emergence of wars and revolutions are all good examples to demonstrate how much burstiness fundamentally characterizes human dynamics. In order to understand the observed bursty phenomena, several modeling approaches have been proposed, assuming various underlying mechanisms such as task prioritization~\cite{Barabasi2005Origin}, reinforcement~\cite{Karsai2012Universal}, memory-driven processes~\cite{Vazquez2007Impact}, and external factors~\cite{Malmgren2008Poissonian}. Moreover, bursty temporal patterns may have relevance in resource allocation design, or could lead to important consequences when interacting with co-evolving dynamical processes. For example, in the case of information spreading on temporal networks, the bursty dynamics of time-varying interactions may slow down spreading~\cite{Karsai2011Small, Miritello2011Dynamical}, thus leading to extremely long times to reach everyone in the network.

In this Chapter we provide a short summary of the most fundamental empirical findings in Section~\ref{sec:obs}, methods and measures in Section~\ref{sec:meas}, and conventional modeling paradigms in Section~\ref{sec:models} of bursty phenomena observed in individual human dynamics. This is only a concise summary of a larger field, thus we would refer the readers, who are interested in studying burstiness, to a recent book on \emph{Bursty Human Dynamics}~\cite{Karsai2018Bursty}, which summarizes the research landscape in more details.

\section{Empirical Findings of Bursty Human Phenomena}
\label{sec:obs}

\begin{figure*}[!t]
\centering
\includegraphics[width=0.85\textwidth]{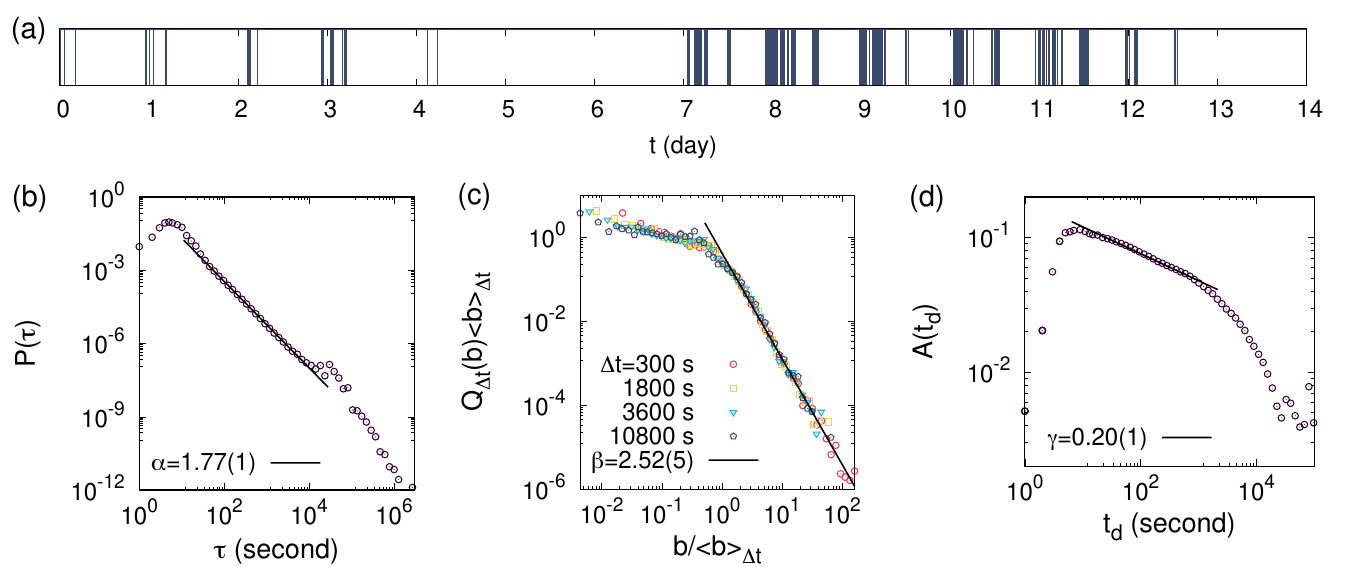}
\caption{Bursty behavior in an event sequence of $n\approx 1.1\times 10^6$ events by one of the most active English Wikipedia editors~\cite{Jo2020Bursttree}. In panel (a) we show a part of the event sequence for the first two weeks since the first edit, where each vertical line denotes an editing event. From the entire event sequence we calculate an interevent time distribution $P(\tau)$ (b), bursty train size distributions $Q_{\Delta t}(b)$ for time windows $\Delta t=300$, $1800$, $3600$, and $10800$ seconds (c), where $\langle b\rangle_{\Delta t}$ is the average burst size for a given $\Delta t$, and the autocorrelation function $A(t_{\rm d})$ as a function of time delay $t_{\rm d}$ (d). In addition, we find the burstiness parameter $B\approx 0.914$ and memory coefficient $M\approx 0.017$.}
\label{fig1}
\end{figure*}

As mentioned, bursty behavior in temporal patterns or time series indicates the presence of a number of events occurring in short active periods separated by long inactive periods~\cite{Barabasi2005Origin}. Such bursty temporal behavior has been observed in a variety of human social phenomena. Bursty human phenomena can be grouped into two main categories, namely, individual-driven and interaction-driven~\cite{Karsai2018Bursty}. Note that these two categories must not be considered exclusive because human individuals are inherently social beings. 

Individual-driven bursty phenomena indicate some activities that are not necessarily concerned with direct interactions between individuals. Examples include job submissions to supercomputers, print requests to the printer, and library loans by the university faculty among many others. In such cases each action, such as a job submission, has been recorded with its timing. A sequence of actions' timings, namely an event sequence, is analyzed using various methods and measures, which will be introduced in the next section.

Interaction-driven bursty phenomena can be grouped into several subcategories mainly according to how individuals interact with each other, more precisely, depending on which communication channel is used for the interaction. The communication channels include face-to-face interactions, mobile-phone based interactions, posted letters and emails, and web-based interactions. The face-to-face interaction might be the most direct communication or interaction between individuals as people should be close to each other both in space and time. By using phones, especially mobile phones in recent years, people can communicate with each other no matter how far they are located. Letters and emails have also been used as communication channels for various purposes. More recently, web-based and application-based services have become more popular and important as communication channels like various Social Networking Services. 

Such temporal interaction patterns have been analyzed in a framework of temporal networks~\cite{Holme2012Temporal, Masuda2016Guide}, by which links (edges) between nodes (vertices) are considered existent or active only at the moment of interaction between nodes. A temporal network can be represented in terms of a tuple for the interaction event $(i,j,t)$, meaning that two nodes $i$ and $j$ interact with each other in time $t$. Then one can choose the level of investigation ranging from a single link or node to the entire network. The temporal pattern of an individual link is obtained as the set of timings of events that occurred on the link. In contrast, the temporal pattern of an individual node is a superposition of those of links adjacent to the node, and so forth. In the other limit, one can aggregate temporal patterns of all links in the network to study the network-level temporal pattern. 

Whether the bursty phenomena are individual-driven or interaction-driven, we get the event sequence as a set of event timings, precisely, $\{t_1,\ldots,t_n\}$ for $n$ events. For example, Fig.~\ref{fig1}(a) shows a part of an event sequence of $n\approx 1.1\times 10^6$ editing events by one of the most active English Wikipedia editors~\cite{Jo2020Bursttree}. Equivalently, the event sequence can also be denoted as a time series $x(t)$ that has the value of $1$ at the moment the event occurred, $0$ otherwise. 

\section{Methods and Measures to Study Bursty Phenomena}
\label{sec:meas}

Most empirical time series of our interest tend to be given in a form of a sequence of event times. Here events indicate actions for individual-driven human dynamics and interaction for interaction-driven human dynamics, respectively. Information on events other than timings, such as duration of action/interaction and other attributes of events, are tentatively ignored in this Chapter. We consider the event sequence as a consequence of the temporal point process that occurs either in discrete or in continuous time.

One of the simplest temporal point processes is a Poisson process by which events randomly occur at a constant rate and independently of each other. This process is called Poisson as the statistics of the number of events in a given time period, denoted by $k$, follows the Poisson distribution:
\begin{align}
    P(k)=\frac{\lambda^k e^{-\lambda}}{k!},
\end{align}
where $\lambda$ denotes an average number of events per time interval. Such a memoryless Poisson process has been used as a reference to study the effect of memory in bursty temporal patterns.

Non-Poissonian or bursty event sequences have been extensively studied for the last two decades and temporal correlations in such event sequences have been characterized by various measures and quantities~\cite{Karsai2018Bursty}. The most accepted characteristics are the interevent time distribution, the burstiness parameter, the memory coefficient, the bursty train sizes, and the autocorrelation function. Each of these five measures captures a different aspect of the bursty time series, while they are not independent of each other. Below we describe these measures mainly following Jo \& Hiraoka~\cite{Jo2023Bursty}.

\begin{figure*}[!t]
\centering
\includegraphics[width=0.6\textwidth]{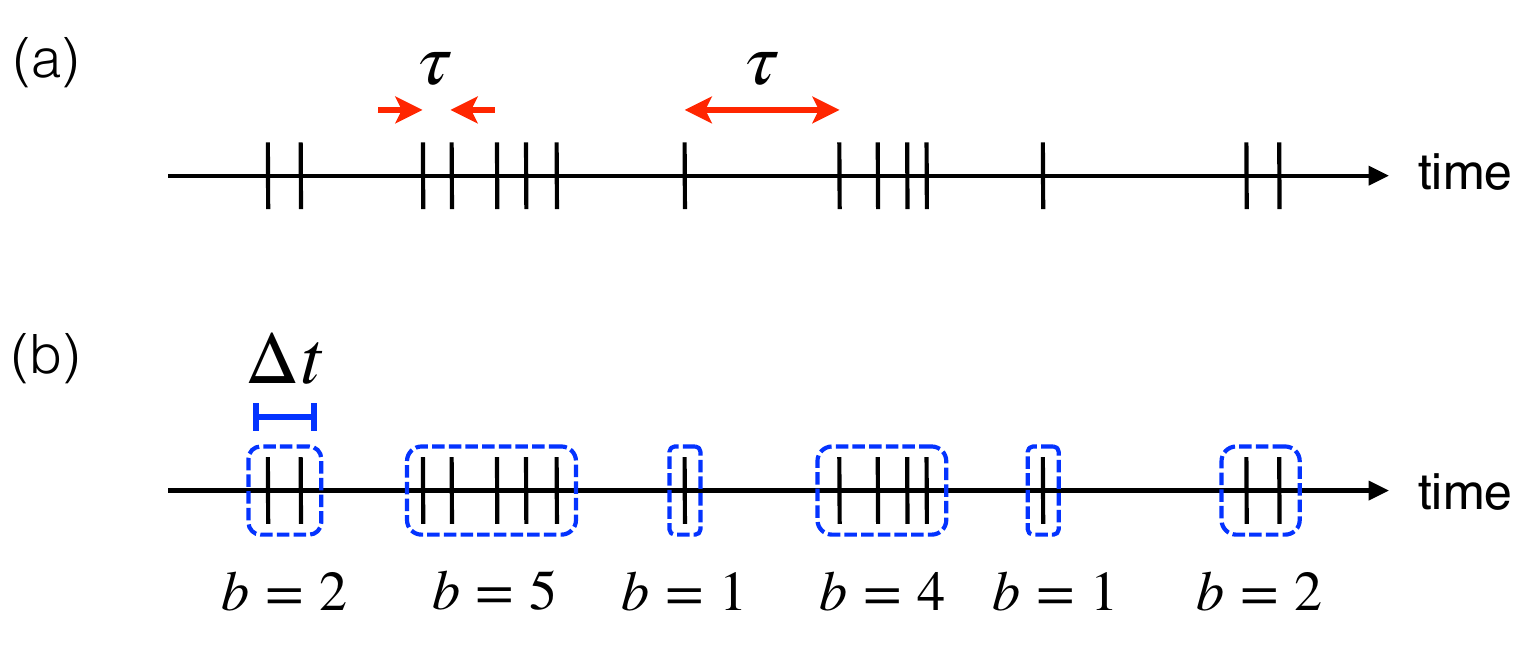}
\caption{Schematic diagrams for bursty time series analysis: (a) definition of the interevent time $\tau$ and (b) definition of the bursty train size $b$ for a given time window $\Delta t$. See the text for detailed definitions.}
\label{fig2}
\end{figure*}

\textbf{Interevent time distribution.} A time interval between two consecutive events in the event sequence defines an interevent time (IET), precisely, $\tau_i\equiv t_i-t_{i-1}$ for $i=2,\ldots,n$, as shown in Fig.~\ref{fig2}(a). For a given event sequence of $n$ events, one can get $n-1$ IETs derived from $n$ events, leading to the IET distribution denoted by $P(\tau)$. The heterogeneous property of IETs is common in bursty time series, and it has often been characterized by the heavy-tailed or power-law IET distribution $P(\tau)$ with a power-law exponent $\alpha$:
\begin{equation}
    \label{eq:Ptau_simple}
    P(\tau)\propto \tau^{-\alpha},
\end{equation}
which may already imply clustered short IETs even with no correlations between IETs. For comparison, the IET distribution obtained from a Poisson process follows an exponential distribution, which is considered to have a thin tail.

\textbf{Burstiness parameter.} The degree of burstiness in the event sequence can be measured by a single value derived from the IET distribution. This single value is the burstiness parameter $B$, and it is defined as~\cite{Goh2008Burstiness}
\begin{equation}
    B \equiv \frac{\sigma - \langle\tau\rangle}{\sigma + \langle\tau\rangle},
\end{equation}
where $\sigma$ and $\langle\tau\rangle$ are the standard deviation and mean of IETs, respectively. For the regular event sequence, all IETs are the same, thus $B=-1$, while for the totally random Poisson process, since $\sigma=\langle\tau\rangle$, one has $B=0$. In the extremely bursty case, as $\sigma\gg \langle\tau\rangle$, one observes $B\to 1$. However, when analyzing the empirical event sequences of finite sizes, the value of $\sigma$ is typically limited by the number of events $n$ such that the maximum value of $\sigma$ turns out to be $\sigma_{\rm max}\simeq \langle\tau\rangle\sqrt{n-1}$, allowing to propose an alternative burstiness measure~\cite{Kim2016Measuring}:
\begin{equation}
    B_n \equiv \frac{\sqrt{n+1}\sigma - \sqrt{n-1}\langle\tau\rangle}{(\sqrt{n+1}-2)\sigma + \sqrt{n-1}\langle\tau\rangle},
\end{equation}
which can have the exact value of $1$ (or $0$) in the extremely bursty case (or in the Poisson process) for any $n$. In turn, this measure does not suffer from finite-size effects.

\textbf{Memory coefficient.} The IET distribution and burstiness parameter characterizing statistical properties of IETs are already quite useful, while they cannot capture correlations between IETs. The memory coefficient can measure such correlations between IETs~\cite{Goh2008Burstiness}. The memory coefficient $M$ is defined as the Pearson correlation coefficient between two consecutive IETs, whose value can be estimated by
\begin{equation}
    M \equiv\frac{1}{n - 1}\sum_{i=1}^{n-1}\frac{(\tau_i - \langle\tau\rangle_1)(\tau_{i+1} - \langle\tau\rangle_2)}{\sigma_1 \sigma_2},
    \label{eq:memory_original}
\end{equation}
where $\langle\tau\rangle_1$ (resp. $\langle\tau\rangle_2$) and $\sigma_1$ (resp. $\sigma_2$) are the average and the standard deviation of the first (resp. last) $n-1$ IETs. Positive $M$ implies that the small (large) IETs tend to be followed by small (large) IETs. Negative $M$ implies the opposite tendency, while $M=0$ is for the uncorrelated IETs. In many empirical analyses, positive $M$ has been observed, while it turns out that $M\approx 0$ for some human dynamics~\cite{Goh2008Burstiness}. 

\textbf{Bursty train size.} Another method for measuring the correlations between IETs is through the analysis of bursty trains~\cite{Karsai2012Universal}. A bursty train is defined as a set of consecutive events such that IETs between any two consecutive events in the bursty train are less than or equal to a given time window $\Delta t$, while those between events in different bursty trains are larger than $\Delta t$, see Fig.~\ref{fig2}(b) for the schematic diagram. The number of events in a bursty train is called bursty train size or burst size, and it is denoted by $b$. The distribution of $b$ would follow an exponential function for any given IET distribution if the IETs are fully uncorrelated with each other. However, in case of correlated events $b$ has been empirically found to be power-law distributed as
\begin{equation}
    \label{eq:burstSizeDistribution}
    Q_{\Delta t}(b)\propto b^{-\beta},
\end{equation}
which has been observed for a wide range of $\Delta t$, e.g., in earthquakes, neuronal activities, and human communication patterns~\cite{Karsai2012Universal}. This indicates the presence of higher-order correlations between IETs beyond the correlations measured by $M$.

\textbf{Autocorrelation function.} The autocorrelation function for an event sequence $x(t)$ is defined with delay time $t_{\rm d}$ as follows:
\begin{equation}
  A(t_{\rm d})\equiv \frac{ \langle x(t)x(t+t_{\rm d})\rangle_t- \langle x(t)\rangle^2_t}{ \langle x(t)^2\rangle_t- \langle x(t)\rangle^2_t},
\end{equation}
where $\langle\cdot\rangle_t$ means a time average. For the event sequences with long-term memory effects, one may find a power-law decaying behavior with a decaying exponent $\gamma$:
\begin{equation}
    A(t_{\rm d})\propto t_{\rm d}^{-\gamma}.
\end{equation}
Temporal correlations measured by $A(t_{\rm d})$ can be understood not only by the heterogeneous IETs but also by correlations between them. In other words the decaying exponent $\gamma$ could be a function of $\alpha$ for the IET distribution and $\beta$ for the burst size distribution. As a special case, when IETs are fully uncorrelated with each other, i.e., for renewal processes, the power spectral density was analytically calculated from power-law IET distributions~\cite{Lowen1993Fractal}. Using this result, one can straightforwardly derive the scaling relation between $\alpha$ and $\gamma$:
\begin{eqnarray}
    \label{eq:alpha_gamma}
    \begin{tabular}{ll}
        $\alpha+\gamma=2$ & for $1<\alpha\leq 2$,\\
        $\alpha-\gamma=2$ & for $2<\alpha\leq 3$.
    \end{tabular}
\end{eqnarray}
More recently, the scaling relation between $\alpha$, $\beta$, and $\gamma$ has been analytically obtained from a discrete-time model considering only one time window $\Delta t=1$~\cite{jo2024temporal}.

The above five measures were used to characterize the event sequence of $n\approx 1.1\times 10^6$ by an individual Wikipedia editor, partly shown in Fig.~\ref{fig1}(a). As a result we observe the heavy-tailed $P(\tau)\propto \tau^{-1.77}$ [Fig.~\ref{fig1}(b)], the heavy-tailed $Q_{\Delta t}(b)\propto b^{-2.52}$ for a wide range of $\Delta t$ [Fig.~\ref{fig1}(c)], and the power-law decaying $A(t_{\rm d})\propto t_{\rm d}^{-0.20}$ [Fig.~\ref{fig1}(d)]. For this event sequence we also find the burstiness parameter $B\approx 0.914$ and the memory coefficient $M\approx 0.017$. For details of the dataset, see Ref.~\cite{Jo2020Bursttree}.

In addition to the above methods, novel methods have been proposed such as a burst-tree representation that can capture the entire temporal correlation in the event sequence by revealing the hierarchical structure of bursty train sizes for the entire range of time windows~\cite{Jo2020Bursttree}. For other metrics, refer to Ref.~\cite{Karsai2018Bursty}.

\section{Modeling Bursty Human Phenomena}
\label{sec:models}

The observation of bursty phenomena in various human dynamical systems called for the identification of underlying mechanisms that induce heterogeneous dynamical activity patterns. Over the last decades several modeling directions have been proposed, based on various concepts of underlying processes, able to explain burstiness at the individual or collective level. Some of these modeling approaches have been successful in reproducing bursty behavior at the phenomenological level, while some others suggested plausible cognitive and social mechanisms, which could explain heterogeneous behavioral patterns. Here we summarize three main modeling paradigms among others~\cite{Karsai2018Bursty}, which gained much attention from the community, each explaining bursty patterns of individual behavior at their simplest definition.

\subsection{Queuing models of burstiness}

The first observations of bursty human phenomena suggested that human activities can be categorized into two classes depending on the IET exponent $\alpha$ in Eq.~\eqref{eq:Ptau_simple}. Letter correspondence and other activities could be characterized by $\alpha=3/2$, while other bursty signals recorded in Web browsing, email correspondence, library loans, etc. were better described by $\alpha=1$~\cite{Barabasi2005Origin, Oliveira2005Human, Vazquez2006Modeling}. Technically, these categories are based on the power-law exponent of the waiting time distribution, denoted by $\alpha_w$, not the power-law exponent of the IET distribution. For this, Barab\'asi~\cite{Barabasi2005Origin} argued that these two exponents could have the same value, i.e., $\alpha=\alpha_w$. These different scaling behaviors could be captured by queuing models, where consecutive rational actions of an individual are assumed to be driven by the execution of prioritized tasks.

For a simple definition~\cite{Cobham1954Priority} let us consider an agent with a priority list of $L$ tasks, each of them assigned with a priority value $x_i$ drawn from a distribution, denoted by $\eta(x)$. The priority values allow the agent to rank the tasks and execute them in a rational order based on their priorities. The quantity of interest is the waiting time $\tau_{\rm w}$ for a task to be spent between its arrival and execution.

One can follow different strategies to update the list of tasks. In the most general case one can assume that priorities are drawn from a homogeneous distribution, i.e., $\eta(x)=1$ for $x\in [0,1]$, and that the priority list can contain an arbitrary number of tasks. Tasks arrive at the rate $\lambda$ following a Poisson process and are executed at the rate $\mu$ by always choosing the one with the highest priority. Solution of this model~\cite{Cobham1954Priority} derives the waiting time distribution:
\begin{equation}
P(\tau_{\rm w}) \propto \tau_{\rm w}^{-3/2} \exp( -\tau_{\rm w} / \tau_0 )
\label{eq:Ptw}
\end{equation}
where $\tau_0\equiv 1/[\mu(1-\sqrt{\rho})^2]$ and $\rho\equiv \lambda/\mu$. If $\rho<1$, tasks are executed right after their arrival, and $P(\tau_{\rm w})$ is reduced to an exponential distribution as $\rho\rightarrow 0$. On the other hand, in the limit $\rho \to 1$ the waiting time distribution appears as a power-law function with exponent $\alpha=3/2$, reproducing well some real-world observations~\cite{Oliveira2005Human}. In this case most of the tasks are executed shortly after their arrival, but some tasks with low priorities may be stuck in the list introducing heterogeneity in the waiting time distribution. Finally, if $\rho>1$, the average queue length is increasing linearly as $\langle l(t) \rangle = (\lambda-\mu)t$, thus a fraction of tasks $1-\rho^{-1}$ will remains in the list forever. Nevertheless, the waiting time distribution of executed tasks still follows the form of Eq.~\eqref{eq:Ptw}.

Motivated by the finite capacity of immediate memory of humans, an extension of this model was proposed~\cite{Barabasi2005Origin}: The model operates with a priority list of fixed size $L$ and it assumes that occasionally tasks with lower priority are performed first. This is introduced by a probability $p$ of the agent to perform the highest priority task in each iteration; otherwise, with probability $1-p$ to select a task randomly from the task list. Thus, in the limit $p\rightarrow 1$ the agent tends to always choose  the task with the highest priority that results in a process with a power-law waiting time distribution with exponent $\alpha=1$ matching the other class of empirical observations reported in \cite{Barabasi2005Origin, Vazquez2006Modeling}. On the other hand, if $p\rightarrow 0$, the agent performs a fully random selection strategy and the waiting times appear to have an exponential distribution.

\subsection{Memory-driven models of burstiness}

Alternative models assuming memory-driven processes were proposed to explain the emergence of bursty phenomena. These models commonly assume non-Markovian correlations between consecutive actions of an individual, modeled by memory functions or reinforcement processes.

One of the first models~\cite{Vazquez2007Impact} introduced a simple memory function to model that subsequent actions of an agent are influenced by its previous mean activity rate. Denoting by $\lambda(t){\rm d}t$ the probability that an agent performs an action within the time interval $[t,t+{\rm d}t]$, one can assume that
\begin{equation}
\lambda(t)=\frac{a}{t}\int_0^t \lambda(t') {\rm d}t',
\label{eq:VazqLamd}	
\end{equation}
where the parameter $a$ determines the degree of dependence on the past activity. The general solution of Eq.~(\ref{eq:VazqLamd}) leads to the indication of a power-law scaling of $\lambda(t)=\lambda_0 a (t/T)^{a-1}$, where $\lambda_0$ is the mean number of actions within the time period $T$. If $a=1$, $\lambda(t)=\lambda_0$ is constant, implying the dynamics follow a Poisson process with an exponential IET distribution. If $a\neq 1$, the process is non-stationary, being either in an acceleration regime ($a>1$) or in a reduction regime ($a<1$). If $a>1$, the system is in the accelerating regime; the IET distribution exhibits a power-law form, $P(\tau)\sim (\tau /\tau_0)^{-\alpha}$, where $\tau_0=1/(a\lambda_0)$ and $\alpha=2+1/(a-1)$ if $\tau_0 \ll \tau < T$. In the reduction regime, if $1/2<a<1$, $P(\tau)$ does not show a power-law behavior, while in the regime $0<a<1/2$ it appears again to follow a power-law with the exponent $\alpha=1-a/(1-a)$ if $\tau \ll \tau_0$.

Another modeling family of correlated burstiness relies on self-exciting stochastic processes of Hawkes type~\cite{Masuda2013SelfExciting}, which concerns the activity rate $\lambda(t)$ defined as follows
\begin{equation}
\lambda(t)=\nu+\sum_{\{i|t_i\leq t\}}\phi(t-t_i),
\end{equation}
where $\nu$ sets the ground activity level, while $\phi(t)$ is called the memory kernel. Several definitions of the memory kernel function have been considered to describe human bursty phenomena. Masuda~et~al.~\cite{Masuda2013SelfExciting} assumed an exponential form $\phi(t)=\alpha e^{-\beta t}$, while Jo~et~al.~\cite{Jo2015Correlated} applied a power-law memory kernel with a finite memory term. Such processes explained the emergence of bursty dynamics in the cases of short-term and long-term memory effects.

Finally, reinforcement mechanisms provide another way to consider memory effects in bursty dynamical processes. Here it is assumed that the longer the system waits since a previous event, the larger the probability that it will keep waiting for the next event. This reinforcement mechanism can induce not only the heterogeneous IETs but also the correlations between IETs commonly observed in real systems. This idea can be demonstrated by a two-state system~\cite{Karsai2012Universal}, which describes an agent who can be either in a normal state $A$ or in an excited state $B$. In a state $A$ the agent performs events at a lower rate, while in a state $B$ events appear at a higher rate. To be precise, the IETs are induced by a reinforcement function in the form of
\begin{equation}
f_{A,B}(\tau)=\left(\frac{\tau}{\tau+1}\right)^{\mu_{A,B}}
\label{eq:KarsaiMemory}
\end{equation}
that gives the probability to wait one time step longer in order to execute the next event after the system has already waited time  $\tau$ since the last event. The exponents $\mu_{A,B}$ control the reinforcement dynamics in states $A$ and $B$, respectively. If $\mu_A \ll \mu_B$, the characteristic IETs in states $A$ and $B$ become fairly different leading to the emergence of temporal heterogeneity in the dynamics. In addition, the state of the system is determined by transition probabilities between two states, leading to long correlated bursty trains in the excited state, separated by long periods of low activities in the normal state.

\subsection{Non-homogeneous Poissonian model of burstiness}

The third modeling paradigm of bursty behavior assumes that human actions are primarily driven by external factors. These external factors introduce a set of distinct characteristic timescales, thereby giving rise to bursty temporal patterns. The most prominent model of such kind aims to describe email correspondence, based on the intuition that human activity patterns are largely determined by circadian and weekly cycles of human individuals and that whenever an event is generated by such cycles, a cascade of activity follows in a short time period~\cite{Malmgren2008Poissonian}. Various timescales involved in the model give rise to heterogeneous temporal behavior with good agreement with empirical observations.

To be precise, the model accounts for periodic activity patterns by using a non-homogeneous Poisson process with a time-varying periodic event rate, i.e., $\rho(t)=\rho(t+W)$, with a given time period $W$ (here it is a week). This rate function captures the convolution of daily and weekly activity distributions of active interval initiation, $p_d(t)$ and $p_w(t)$, as
\begin{equation}
\rho(t)=N_w p_d(t) p_w(t),
\label{eq:NHPoissRate}
\end{equation}
where $N_w$ is a constant representing the average number of active intervals during a week. Each event generated by $\rho(t)$ initiates a secondary process, a cascade of activity, modeled by a homogeneous Poisson process with rate $\rho_a$. During the cascade of activity, $N_a$ additional events, drawn from a distribution $p(N_a)$, occur before being governed by $\rho(t)$ again. Thus this model is fully determined by $N_w$, $p_d(t)$, $p_w(t)$, $\rho_a$, and $p(N_a)$, which turns out to reproduce precisely the observed heterogeneous bursty dynamics in individual email correspondences.

\section{Conclusion}
\label{sec:disc}

In this Chapter, we have presented a brief overview of bursty human phenomena, where to observe them, how to measure them, and what kind of modeling approaches have been provided to capture their origin. We mainly focused on the characterization of human bursty systems at the level of single actions or interactions. Our goal was not to provide a complete review of all results in this field but to indicate a starting point for the interested readers, and to highlight the main research tracks found over the last years.

It is important to mention that bursty human dynamics have indisputable consequences on dynamical processes. The heterogeneous timings of actions and interactions largely control the possible transmission of any kind of information between interacting peers, or the timely connectedness of the underlying temporal structure~\cite{Pastor-Satorras2015Epidemic}.

Although it is somewhat straightforward to find examples for these effects in the larger area of spreading processes, there are several other dynamical processes where these phenomena play role. Random walks, evolutionary games, or opinion formation dynamics are only a few examples, where bursty characters of the interacting agents can critically alter the final outcome of the dynamical process.

We hope that this short introduction to the extensively explored yet not fully understood phenomena of bursty dynamics will motivate readers to research this area more deeply~\cite{Karsai2018Bursty}. This may lead to the exploration of new examples of dynamical systems exhibiting this character, and to deeper and more comprehensive explanations of its reasons and consequences.

\subsection*{Further Reading}
We refer the reader to a recent book \emph{Bursty Human Dynamics}~\cite{Karsai2018Bursty} on the observations, characterization, and modeling of bursty human phenomena.

\begin{acknowledgements}
The authors thank their long-time colleague Prof. Kimmo Kaski at Aalto University, Finland.
\end{acknowledgements}


%

\end{document}